# Evaluation of the impact of Heat-Wave on Distribution System Resilience

Andrea Mazza, Yang Zhang,
Ettore Bompard, Gianfranco Chicco
*Dip. Energia "Galileo Ferraris"*
*Politecnico di Torino*
Torino, Italy
name.surname@polito.it

Emiliano Roggero, Giuliana Galofaro

*IRETI SpA*

Torino, Italy
name.surname@ireti.it

*Abstract*—This paper presents the findings about the impact of heat waves on a real urban distribution system. A data-driven methodology is proposed to simulate the portion of faults that can be associated to normal conditions (and hence to reliability) and the portion correlated to the heat wave occurrence. Based on real data collected in the years 2012-2017, the fault rates associated to reliability and resilience have been calculated and then used to feed a Monte Carlo simulation aiming to manage the uncertainty in the fault occurrence. Finally, based on the Italian legislation, the benefits deriving by the substitution of the faulted portion of the system have been calculated.

## I. Introduction

As the backbone of modern industrialized society and economics, a highly available power system supply plays a significant role in people's daily life [1]. However, with the global climate change, electrical infrastructures are exposed to a harsher environment all over the world [2]. In recent years, many power outages triggered by the natural hazards have occurred. For example, over 500,000 Long Island Power Authority customers lost power in 2011 after being hit by the hurricane Irene [3]. It took 8 days before the restoration of 99% of the company's customers. The total number of customers affected by this hurricane is more than 4.3 million people on the East Coast of the US.

In case of lack of proper supply, Distribution System Operators (DSOs) need to pay penalties to the Regulatory Authority (excluding the causes depending on the final user or classified as *force majeure* – including catastrophic events). Therefore, both sides (customers and operators) requires more reliable power system also in case of adverse environmental conditions. It has been reported that, in Europe, the share of investments in distribution network with respect to the total on power grids is supposed to continuously grow from 66% in 2020 to 80% by 2050 [4]. An important part of those investments will address urban distribution systems, due to the increasing share of people living in urban areas (approaching 68% of the total population in 2050 [5]).

In addition to the reliable operation of the grid under common weather conditions, the ability to withstand extraordinary and high-impact low-probability (HILP) events is also needed in the planning and operation of modern power systems [2], because the rare events could cause huge damages to the fundamental infrastructure and have a tremendous social impact. Therefore, the concept of *resilience*, as the attitude of the system to withstand extreme events, has become an emerging topic in power system analysis [10]. A probabilistic methodology is proposed in [11] to evaluate the adaptation measures to increase the resilience of power system to natural disasters. This method is also capable to deal with the multi-hazard and multi-risk analysis power system resilience. A multi-phase resilience assessment framework is developed in [12], which is used to analyze the natural threat on critical infrastructures. Different strategies to boost the resilience of power systems are also discussed with the multi-phase adaptation cases. The effects of catastrophic weather on power system is evaluated with a fuzzy clustering method in [6]. The cascading failures in power grid, caused by natural hazards, are analyzed with an extreme weather stochastic model in [7]. In Italy, with the heat waves in urban areas, increasing faults have been recorded in distribution network in a relatively concentrated period [9]. These dense occurrences of faults bring a severe threat to the secure operation of the urban distribution system. Due to the structure of the distribution system, which is a weakly-meshed system operated in radial configuration, the single fault on the feeder can be isolated without losing any customers for a long time. However, the heat wave could cause successive faults happening on the same feeder in a relatively short time from the first one, i.e., the successive fault happens before the first one is repaired. This leads to severe blackouts for the customers between the two fault locations. This is a typical extreme weather-related power interruption with high impact and low probability.

From the planning point of view [13], Cost Benefit Analysis (CBA) has to be applied to evaluate the improvement of the resilience of the network consequent to certain investments. However, the low probability of extreme weathers brings challenges that, summed up to the uncertainties of occurrence of natural disasters, requires an appropriate methodology.

This paper considers heat waves in an urban area of Italy with HILP events hitting the local distribution system. The occurrence probability of heat waves is discussed on the basis of 6 year-long weather records. On the basis of the fault records provided by the local DSO, the fault rates for single and repetitive faults (both in the presence and absence of heat waves) are calculated and used in a Monte Carlo simulation model to evaluate, over a 25-year time horizon, the benefits obtained after the upgrade of the underground cables.

This paper is structured as follows. Section II presents the methodological framework, which can be used for any HILP threats. In Section III, the phenomenon of heat wave in an Italian urban area is presented, by considering the data collected in the urban area of the city of Turin (Italy). The concept of repetitive faults is then introduced, by presenting how they are more likely to happen in presence of heat waves and can strongly affect the supply quality for the users. The CBA of the investment in the distribution network is detailed in Section IV and the results on different real cables are shown in Section V. Finally, Section VI reports the concluding remarks.

## II. METHODOLOGY

The general framework for incorporating the calculation of the CBA within a grid resilience evaluation is shown in Fig. 1.

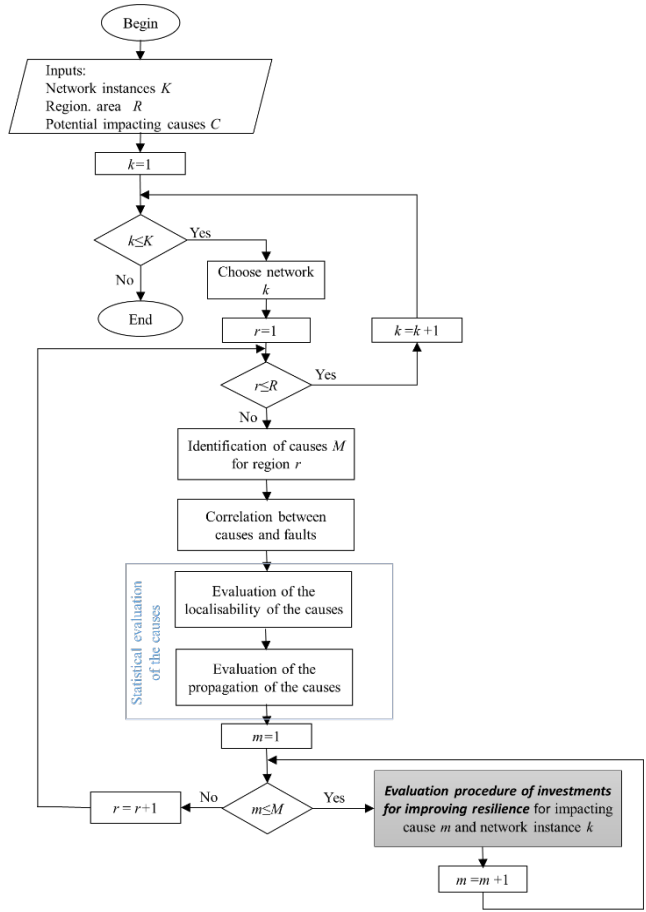

Fig. 1. Flow chart of the general approach for the resilience evaluation including CBA.

The procedures require different steps, as follows:

*1) Collection of network instances:* the criterion for the choice of the network instances to be collected is related to the level of similarity of the portion of network considered with respect to the distribution networks operated by the same DSO, or simply based on network portions that experienced issues in case of extreme weather conditions. As reported in Fig. 1, the number of network instances is named as $K$.

*2) Enumeration of the possible causes and filtering:* this part is necessary if the calculation is made on different regions, with different causes (e.g., in case of a unique DSO operating in wide area).

*3) Correlation between causes and faults:* it is possible to correlate statistically the causes and the number of faults in the network (measured with one of the usual metrics, such as Energy Not Supplied, Number Of Customer Unsupplied, Duration of the Lack of Supply and so on). This correlation means also to recognize the level of the stress coming from the causes which can lead to have a fault in the network.

*4) Evaluation of the localisability of the causes*: this is a concept connected to what is considered as "minimum" component of the system under analysis. In fact, for a matter of example, the occurrence of a flood in a certain region implies to consider the entire system, and not the single component. However, if a cloud burst occurs, the aggregation of components should be considered: the aggregation is composed of all the components subject to faults due to the cause. In case of snow/icing, the interest is on the single most representative length of a line, and thus a component-based evaluation has to be carried out.

*5) Evaluation of the propagation of the causes:* in case of an extreme weather event, the cause can propagate towards close regions. For this reason, it is necessary to take into account this possibility, for considering also multiple events due to the same cause.

*6) Define the vulnerability of the components/system to the causes:* in this case, the vulnerability of the component/system is linked to the different causes and to the value of the stress variables.

*7) List of the possible remedial action:* by knowing the faults happened so far and the causes, a list of investments can be made.

*8) Evaluation of the investments by CBA:* this point depends on the regulatory framework in which it is developed. In case of output-based regulation, the improvement in terms of both reliability and resilience can be considered.

If more than one investment strategy are evaluated, a Decision making procedure, which can be multicriteria, can be used, to rank the different investments on the basis of the CBA's outputs.

## III. HEAT WAVE PHENOMENON

### A. Theoretical framework

The heat wave phenomena are usually defined as periods with exceptionally hot weather hitting portions of territory that, usually, are not used to them. For a matter of nomenclature, the *heat wave* happens in summer, whereas hot periods in winter are indicated as *warm spells* [14]. However, it is necessary to translate the qualitative description into appropriate indices that quantify the presence of this kind of phenomenon. In [15] an index named Excess Heat Factor (EHF), originally developed for describing heat wave phenomena in Australia [16], has been applied to describe the presence of heat waves in Greece. The same index has been applied to describe the occurrence of heat waves in Czech Republic [17] and Romania [18]. Due to the wide use of this index in different contexts (both on the seaside and internal territories) and based also on the fact that Greece and Italy are lying within the Mediterranean Basin, which recently has seen an increase of heat wave phenomena (see for example [19]), this index has also been used to describe the occurrence of heat waves in Turin (Italy). The EHF index refers to single days and combines both the historical shape of temperatures and the effect on humans, i.e., the *long-term* temperature drift and the *short-term* temperature drift effect, as reported in Eq. (1):

$$EHF_i = EHI_{sig,i} \cdot max\{1, EHI_{accl,i}\} \quad (1)$$

where $EHI_{sig}$ is the *significance index* and $EHI_{accl}$ the *acclimatation index*. The first one aims to measure the deviation from the historical conditions, whereas the second one evaluates the impact of short-term and sharp temperature variations. Their definitions are shown in Eqs. (2) and (3):

$$EHI_{sig,i} = \frac{(T_i + T_{i-1} + T_{i-2})}{3} - T_{95} \quad (2)$$

$$EHI_{accl,i} = \frac{(T_i + T_{i-1} + T_{i-2})}{3} - \frac{\sum_{k=3}^{32} T_{i-k}}{30} \quad (3)$$

where both terms are defined by considering the average temperature of the day under analysis and two days ahead. However, $EHI_{sig,i}$ considers as reference the 95[th] percentile of the daily average temperature ($T_{95}$, calculated over a period of at least 30 years), whereas $EHI_{accl,i}$ considers as reference temperature the average value calculated over the past 30 days. Note that the original formulation of [14],[16] has been slightly modified in [17], in particular to calculate of the average temperature on the three days, and this formulation has been adopted also for coherence with [15].

### B. Calculation for the city of Turin

The definition of *EHF* has been applied by considering the data available on the Piedmont Agency for the Environment [20]. The value $T_{95}$ has been computed by considering the daily temperature over the period 1989–2011. Thanks to this historical information, the occurrence of heat waves in the period 2012–2017 have been evaluated, as shown in Fig. 2.

It is evident that the year 2017, in terms of number of heat waves, was the worst one (with six heat waves), followed by the year 2016, with four heat waves. Beyond the number of heat wave occurrences, it is interesting to evaluate the number of days per each occurrence, shown in Fig. 3. The longest heat wave occurrence happened in 2015, reaching 30 days of duration. This record is followed by 2017, when the longest heat wave period reached 15 days.

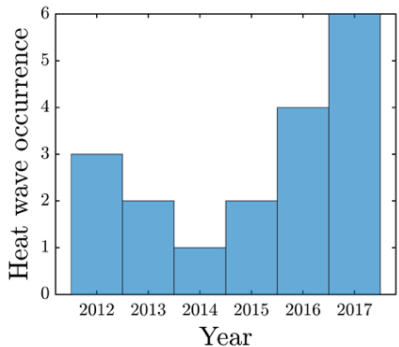

Fig. 2. Heat wave occurrence in the period 2012 – 2017.

From the above data it is possible to obtain the following information:
- The total number of heat waves happening between 2012 and 2017 is 18.
- The average value of heat waves is 3 per year.
- The average duration of each heat wave is 7.17 days.
- The minimum time between one heat wave and the following one is three days.

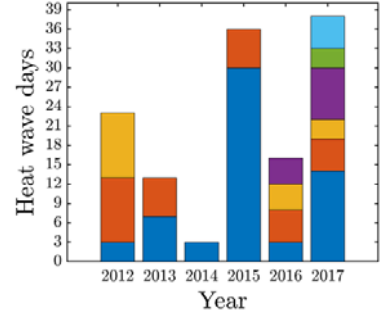

Fig. 3. Duration of every heat wave occurrence in the period 2012–2017.

The same Environmental Agency provides the expected heat waves that will affect the City of Turin in the next years. They have been extrapolated by using the indications of the International Panel for the Climate Change (IPCC). In particular, the values have been obtained by considering the IPCC scenario named RCP4.5 (see [21][22]). As seen in Table I, the heat wave phenomena will become tougher in terms of *total number of days* and *maximum duration of the event*, and this highlights one more time the need to have a proper tool to face the issues connected with the heat wave occurrences.

## IV. DISTRIBUTION SYSTEM FAULTS

### A. Data Analysis

The analysis of the distribution system faults have been carried out on a database referring to the Turin's urban distribution system (operated by IRETI SpA) and reporting all the faults registered in the period 2012–2017. The DSO registered an increasing number of faults that affected underground cables (and their joints in particular) during the summer 2017. This fact suggested to limit the database by considering only the faults that refer to portions of underground cables, being more sensitive to the temperature increase (as also recently reported in [23]). Globally, 1042 faults have been registered: the faults have been categorized into *single faults* and *repetitive faults*. The single faults can be defined as permanent faults that hit a single component of the feeder and are not followed by any other fault occurrence.

The single faults are more common and, once located, the customer supply can be restored by exploiting the weakly-meshed structure of the distribution system. In the analysed database, the number of faults falling in this category was 756 (about 73%).

The repetitive faults category, instead, includes all the faults that hit the feeders derived from the same HV/MV substation within a defined time interval $T_f$. Even though these faults are less common, they can affect more considerably the quality of the service, because multiple contingencies have to be faced in a short time frame. Practically, the time interval $T_f$ represents the time required to fix a fault: if another fault occurs in that time interval, hence the single fault event becomes a repetitive fault event. In the analysed fault database, the number of faults composing the group of repetitive faults was 286 (around 27% of the total). For the sake of simplicity, and in particular because their number was very low, the cases counting more than two repetitive faults have been neglected. For this reason, the number of repetitive fault occurrence (composed by at least two faults) is 119.

From the practical point of view, we can imagine to sweep the fault occurrence to find all the faults that may be single, i.e., occurring after a time interval higher than $T_f$ with respect to the previous one. Then, the second sweep concerns the remaining faults associated to their initial faults (seen as a center of an arc circle), as visualized in Fig. 4.

The example considers $T_f = 8$ h, and it is evident that fault 4 and fault 5 may be associated to fault 3, because happening in the same arc of interest within the time interval $T_f$.

Different time intervals $T_f$ have to be defined, according to the period in which the faults happen. In fact, the repetitive faults can be further classified as either *faults affecting the reliability* of the system or *faults affecting the resilience* of the system. The difference lies into the presence of the heat wave phenomena. Beyond the time between the faults $T_f$, other time intervals are important for defining the effect of the faults on the user and DSO sides, i.e., the fault location duration $T_{FL}$, and the average time in which the customer experiences the lack of supply in case of permanent fault $T_{UN}$. It is worth noting that all the values reported in Table II have been obtained from the local DSO and, only for the $T_f$ referring to the resilience, by the Italian Regulatory Framework [24]. The difference in the $T_f$ values aims to represent that it may be easier to fix faults when the weather conditions are not exceptional.

TABLE I. HEAT WAVE PHENOMENA 2022–2046

| Y. | No. Heat Waves | Longest Heat Wave Days | Total Number of Heat Wave days | Y. | No. Heat Waves | Longest Heat Wave Days | Total Number of Heat Wave days |
|---|---|---|---|---|---|---|---|
| 2022 | 3 | 13 | 23 | 2035 | 2 | 8 | 11 |
| 2023 | 1 | 8 | 8 | 2036 | 3 | 27 | 45 |
| 2024 | 1 | 9 | 9 | 2037 | 1 | 10 | 10 |
| 2025 | 2 | 11 | 14 | 2038 | 2 | 5 | 10 |
| 2026 | 2 | 3 | 6 | 2039 | 1 | 6 | 6 |
| 2027 | 2 | 22 | 34 | 2040 | 0 | 0 | 0 |
| 2028 | 4 | 7 | 19 | 2041 | 6 | 12 | 38 |
| 2029 | 1 | 3 | 3 | 2042 | 5 | 9 | 27 |
| 2030 | 3 | 13 | 32 | 2043 | 4 | 9 | 23 |
| 2031 | 5 | 9 | 28 | 2044 | 2 | 9 | 14 |
| 2032 | 3 | 14 | 25 | 2045 | 4 | 8 | 24 |
| 2033 | 2 | 3 | 6 | 2046 | 4 | 11 | 26 |
| 2034 | 3 | 5 | 14 | - | - | - | - |

The occurrence of a new fault within the time interval $T_f$ shifts the fault type from single to repetitive. However, the time required for the fault location and the time when the users experience the lack of supply cannot be defined *a priori* but depend on when the second fault occurs. From the analysis of the fault database, it was possible to get the time of occurrence of the second faults $T_{SF}$, both with and without heat waves, as reported in Table III.

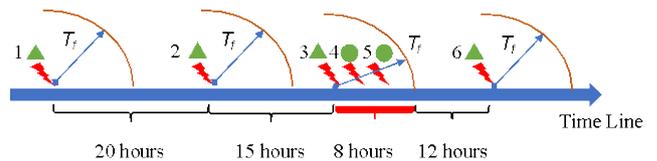

Fig. 4. Visual representation of the repetitive fault grouping procedure.

TABLE II. DIFFERENT TIME INTERVALS FOR FAULT CALCULATION

| Times | Type of faults | | |
|---|---|---|---|
| | *Single faults* | *Repetitive Faults (Reliability)* | *Repetitive Faults (Resilience)* |
| $T_f$ | - | 6 h | 8 h |
| $T_{FL}$ | 1 h | > 1 h | > 1 h |
| $T_{UN}$ | 40 min | Varying | Varying |

TABLE III. DIFFERENT TIME INTERVALS FOR FAULT CALCULATION

| Occurrence of the second fault w.r.t. the first one | Reliability | Resilience |
|---|---|---|
| $T_{SF} \leq 1$ h | 85% | 86% |
| 1 h < $T_{SF} \leq 2$ h | 8% | 10% |
| 2 h < $T_{SF} \leq 3$ h | 1% | 4% |
| 3 h < $T_{SF} \leq 4$ h | 2% | 0 |
| 4 h < $T_{SF} \leq 5$ h | 4% | 0 |
| 5 h < $T_{SF} \leq 6$ h | - | 0 |
| 6 h < $T_{SF} \leq 7$ h | - | 0 |

The values show that, in case of heat wave, the repetitive fault happens within 3 h from the initial one, whereas without heat wave the repetitive faults may happen up to 5 h after the initial fault. Starting from the value of fault, is possible to evaluate the fault rate for any type of fault, in different weather conditions. Due to the fact that heat waves happen in summer and their duration may change over the year, the fault rate (at system level) have been calculated at *daily* granularity instead of yearly granularity. Three different types of days are considered:

- Summer days when the heat wave occurs, denoted as $d_s^{(hw)}$
- Summer days when the heat wave does not occur, denoted as $d_s$
- All the other days, denoted as $d_w$

Hence, it is possible to calculate three different fault rates as follows:

$$\lambda_s^{(hw)} = \frac{\sum_{y=1}^{Y} F_s^{(hw)}(y)}{L \cdot \sum_{y=1}^{Y} d_s^{(hw)}(y)} \quad (4)$$

$$\lambda_s = \frac{\sum_{y=1}^{Y} F_s(y)}{L \cdot \sum_{y=1}^{Y} d_s(y)} \quad (5)$$

$$\lambda_w = \frac{\sum_{y=1}^{Y} F_w(y)}{L \cdot \sum_{y=1}^{Y} d_w(y)} \quad (6)$$

where $Y$ indicates the number of years under analysis (in our case 6 years, corresponding to the period 2012-2017), $L$ is the total length of the distribution system under analysis (about 2000 km for the Turin's system [25]), $F_s^{(hw)}$ is the number of faults occurring during the heat wave phenomena, $F_s$ is the number of faults occurring during the summer days when no heat waves occur, whereas $F_w$ indicates the number of faults occurring in the other days. Due to the nature of the heat wave phenomenon, the "summer" period has been considered to be lying between 1st May and 30th September. The values of the fault rates for single and repetitive faults are shown in Table

IV, where it is possible to see that:
- The existence of heat waves affects both the single fault and the repetitive fault occurrences.
- In summer, if no heat wave occurs the repetitive fault rate is almost four times higher than in winter, but about one fourth with respect to the one calculated when the heat wave occurs.
- The fault rates in days which do not belong to the summer period have the lowest values.

TABLE IV. FAULT RATE VALUES

| Fault rates (fault·km$^{-1}$·day$^{-1}$) | Single Fault | Repetitive Fault |
|---|---|---|
| $\lambda_s^{(hw)}$ | 0.00022 | 0.000054 |
| $\lambda_s$ | 0.00022 | 0.000042 |
| $\lambda_w$ | 0.00014 | 0.000015 |

*B. Fault simulation approach*

The faults have been modelled by using the Poisson process, which properly models the rare event occurrences. The Poisson process is based on three hypotheses:
- The number of events, at the beginning of the period under analysis, is null.
- The event occurrences are *independent* of each other.
- In any interval with duration *t*, the number of events can be represented through a Poisson distribution with mean value linked to the duration *t*

It is worth nothing that the approach used in this paper meets all the criteria, because (*i*) the analysis starts from a healthy grid, (*ii*) the approach considers single and repetitive faults, and these ones group multiple faults (with common cause "heat waves") within a unique fault event characterized by a unique fault rate, and (*iii*) the occurrence of a fault is a *rare event*, properly modelled by Poisson distribution.

V. DEFINITION OF THE BENEFITS

In real systems the evaluation of the benefits consequent to any network investment cannot be really decoupled between reliability and resilience. As matter of the fact, the reduction of the number of faults is simply seen as an improvement of the network performance, without any consideration on what caused it. However, in some legal frameworks (as in Italy [24]), it is important to model separately the resilience improvement of the network and the reliability of the network. In particular, it is possible to recognize four different benefits:
- *Benefit B1*, referring to the reduction of the time during which the customers are unsupplied thanks to the decrease of the faults in presence of heat wave.
- *Benefit B2*, in terms of lower cost for fault location and restoration, that the DSO has owing to the reduction of faults in presence of heat wave.
- *Benefit B3*, referring to the reduction of the time during which the users are unsupplied thanks to the decrease of the faults when no heat wave occurs.
- *Benefit B4*, in terms of lower cost for fault location and restoration, that the DSO has owing to the reduction of faults when no heat wave occurs.

It is worth noting that *B1* and *B3* refer to the system (i.e., are system performance indicator), whereas *B2* and *B4* are considered as direct benefits of the DSO.

The faults taken into account for the calculation of *B4* are solely the *repetitive faults* during heat wave phenomena, whereas all the other types of faults (i.e., single faults in all the periods and repetitive faults in winter and in summer when there is no heat wave occurrence) are considered as faults affecting the reliability.

VI. CASE STUDY

The case study considers a real portion of the urban distribution system of the city of Turin, whose schematic is shown in Fig. 5. It is composed of 21 MV/LV substations, spread on three feeders. One of the feeders is connected to the HV/MV substation, whereas the other two are connected to other system portions, to guarantee an alternative supply path in case of permanent faults. The total active power of the grid portion is 11 MW for residential load and 6 MW of non-residential load.

According the current Italian regulation, the *ENS* is monetized 54 €/kWh for non-residential customers and 12 €/kWh for residential customers. Following the indication of the DSO, the hourly cost for the fault location team is 95 €/h, whereas the hourly cost for the team for service restoration is 250 €/h. The cost to rent the mobile generation is 1000 €/day.

The case study considers three different scenarios, based on the reduction of the fault rate reached thanks to the investment, i.e., 20%, 50% and 80%. This information may be obtained through an extensive test campaign reproducing the heat wave conditions using both an old and a new cable (and related joints).

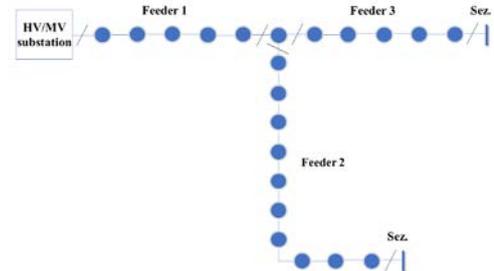

Fig. 5. Schematic of the grid portion under analysis.

As shown in Fig. 6 and Fig. 7, the benefits related to the reliability are much higher than the benefit of the resilience. Furthermore, comparing the two figures, it is evident that the benefit for the users (in terms of improvement of the quality of the service) is higher than the own benefits of the DSO (measured in terms of lower fault-related costs).

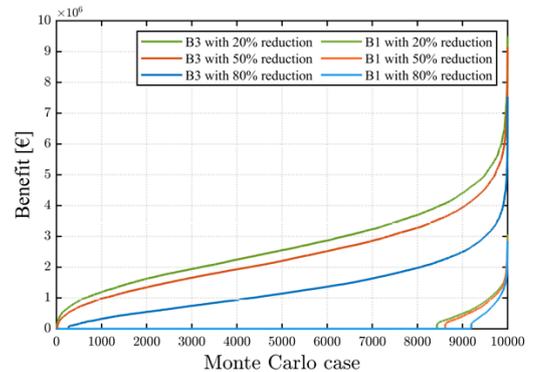

Fig. 6. Benefits B1 and B3 (customer side).

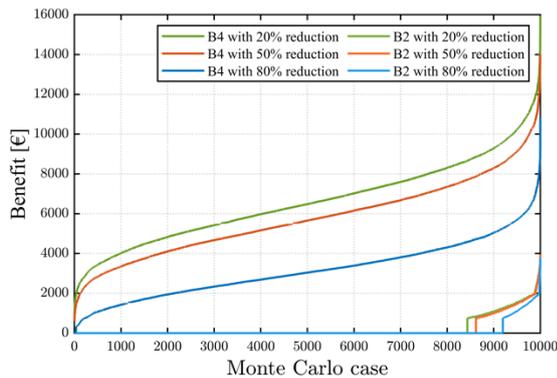

Fig. 7. Benefits B2 and B4 (DSO side).

Furthermore, it is interesting to see the different *shapes* of the benefit terms related to reliability and resilience: while the reliability terms are practically always different from zero and have a quite constant growth (except for the right-hand side of the curve), the resilience terms are mostly zero (low probability), with an initial step (i.e., in the presence of extreme weather the faults create an actual damage) and with a sharper growth (especially in the last part of the curve).

All the above points suggest that the *average value* may be meaningful for describing the reliability indicator but is totally meaningless for resilience. Hence, the Regulatory bodies should consider this aspect, especially in case of *output-based* regulation aiming to remunerate the DSO investment in the resilience improvement.

## VII. CONCLUSIONS

This paper focused on resilience evaluation in distribution systems. After the introduction of a general framework for the study of the resilience, an introduction of the heat wave phenomena has been provided. Their occurrence in the territory of the City of Turin has been demonstrated through the calculation of a proper indicator based on historical data. Then, starting from a database, the faults have been divided into *single* and *repetitive*, and their occurrence with and without heat wave have been simulated for evaluating the impact of the investment on four benefits defined by the Italian Regulatory body. The benefit evaluation considers three different reduction factors, providing a sensitivity analysis. Future works in this area will consider new elements for the evaluation of the post-investment condition, but also new threats, for providing a set of solutions that can be offered to different system operators.